\documentclass[hyper]{JHEP} 
\usepackage{epsfig}
\newcommand\fverb{\setbox\pippobox=\hbox\bgroup\verb}
\newcommand\fverbdo{\egroup\medskip\noindent%
            \fbox{\unhbox\pippobox}\ }

\newcommand\fverbit{\egroup\item[\fbox{\unhbox\pippobox}]}
\newbox\pippobox

\title{Note About String with Euclidean World-Sheet
in $AdS_5\times S^5$}
\preprint{\hepth{0802.0935}}

\author{by J. Kluso\v{n}\\
     Department of Theoretical Physics and Astrophysics\\
                   Faculty of Science, Masaryk University\\
Kotl\'{a}\v{r}sk\'{a} 2, 611 37, Brno\\
Czech Republic\\
    E-mail: \email{klu@physics.muni.cz}}
\abstract{ This note is devoted to the
study of the classical solutions on the
bosonic string with euclidean
world-sheet in $AdS_5\times S^5$. We
generalize solutions presented in
arXiv:0707.4254 [hep-th] to the case
 where we include the dynamics of the string  on
$S^5$ as well.}

\keywords{AdS/CFT correspondence}

\def\tgamma{\tilde{\gamma}}

\newcommand{\tphi}{\tilde{\phi}}

\newcommand{\talpha}{\tilde{\alpha}}
\newcommand{\tbeta}{\tilde{\beta}}

\newcommand{\tomega}{\tilde{\omega}}

\newcommand{\tLambda}{\tilde{\Lambda}}
\begin{document}
\section{Introduction}
The $AdS/CFT$ correspondence
\cite{Maldacena:1997re,Witten:1998qj,Gubser:1998bc}
\footnote{For review, see
\cite{Nastase:2007kj,Klebanov:2000me,D'Hoker:2002aw,
Klebanov:2000me,D'Hoker:2002aw}.}
 shows deep
relations between the $N=4$ super
Yang-Mills (SYM) theory and the string
theory in $AdS_5\times S^5$ where the
classical string solutions play an
important role \cite{Gubser:2002tv},
for review and extensive list of
references, see
\cite{Tseytlin:2004xa,Tseytlin:2003ii}.
The energies of classical strings have
been shown to match with the anomalous
dimensions of the gauge invariant
operators while an open string that
ends on a curve at the boundary of
$AdS_5$ has been analyzed to study the
strong coupling behavior of the Wilson
loop in the gauge theory
\cite{Rey:1998ik,Maldacena:1998im,Drukker:1999zq}.

Recently
 Alday and Maldacena in
 remarkable paper
 \cite{Alday:2007hr}
\footnote{For some recent papers, see
\cite{Drummond:2007bm,Itoyama:2007fs,Itoyama:2007fs,
Drummond:2007au,Jevicki:2007aa,Itoyama:2007ue,Yang:2007cm,
Mironov:2007xp,Ricci:2007eq,Popolitov:2007wc,
Astefanesei:2007bk,Ryang:2007bc,Alday:2007he,McGreevy:2007kt,
Drummond:2007cf,Roiban:2007dq,Mironov:2007qq,Alday:2007mf,
Kruczenski:2007cy,Brandhuber:2007yx,Drummond:2007aua,
Buchbinder:2007hm,Abel:2007mw}.}
computed the planar $4$-gluon
scattering amplitude at strong coupling
in the $N=4$ SYM theory using AdS/CFT
correspondence.  The $4$-gluon
scattering amplitude was evaluated
as the string theory computation of the
$4$-cusp Wilson loop composed of $4$
light-like segments in the T-dual
coordinates where a certain open string
solution in $AdS_5$ space is found to
minimize the area of the string surface
whose boundary conditions are
determined by the massless gluon
momenta and a dimensional regulariztion
is used to regularize the IR
divergence.

As was shown in
\cite{Kruczenski:2007cy} these results
are closely related to the remarkable
observations in perturbative (planar
$N=4$) gauge theory: The scaling
function $f(\lambda)$ can be either
found as a coefficient in the anomalous
dimension of gauge invariant large spin
twist two operator or as a cusp anomaly
of a light-like Wilson line
\cite{Bassetto:1993xd,Korchemsky:1992xv}.
Then it was  shown in
\cite{Kruczenski:2007cy} that this fact
has nice explanation in dual
perturbative $AdS_5\times S^5$ where
the anomalous dimension of minimal
twist operator is represented either by
the energy of a \emph{closed} string
with large spin $S\gg 1$ in $AdS_5$
\cite{Gubser:2002tv} or it follows the
\emph{open} string picture, i.e. from
the area of a surface ending on a cusp
formed by two light-like Wilson lines
on the boundary of $AdS_5$ 
\cite{Kruczenski:2002fb}. Then it was
argued that these two approaches are
(under specific scaling limits) close
related when they become equivalent
upon certain analytic continuation that
is needed  to convert the Minkowski
world-sheet coordinates in the closed
string case into the euclidean one in
the open string Wilson loop case and
$AdS_5$, i.e. conformal $SO(2,4)$
transformation. Then it was shown in
\cite{Kruczenski:2007cy} that the
world-sheet surface studied in
\cite{Alday:2007hr} can be related
(before an IR regularization) to the
cusp Wilson loop surface found in
\cite{Kruczenski:2002fb} using
$SO(2,4)$ isometry of $AdS_5$.

It is remarkable  that  string with
euclidean world-sheet that is embedded
in ordinary Minkowski space-time plays
such a cruical role in AdS/CFT
correspondence. In fact, it is well
known that string theories naturally
contain in their spectra  extended
objects with euclidean world-sheet
signature (S-branes, S-strings
\cite{Gutperle:2002ai,Lambert:2003zr,Hashimoto:2003qx})
even  if their   precise definition is
unclear. On the other hand we mean that
it is certainly important
 to study   properties of these
objects  and try to identify their
possible applications.  In fact the
goal of this paper is to investigate
the dynamics of the bosonic string with
the euclidean world-sheet metric  in
$AdS_5\times S^5$ and try to see how it
is possible to extend the classical
solution found in
\cite{Kruczenski:2007cy} to more
general case. Recall that the ansatz
given in \cite{Kruczenski:2007cy}
describes light-like Wilson line that
ends on the boundary of $AdS_5$. Our
goal is to generalize this solution to
the case when we allow non-trivial
configuration of the string on $S^5$. We
find that in case of light-like
Wilson line solution
 the dynamics of the string on
$AdS_5$ decouples from the dynamics on
$S^5$ as a consequence
 of the fact that $AdS_5$ part of the
 Virasoro constraints vanishes
 separately. On the other hand
it is important to stress
that   solution that describes
 dynamics of the string on $AdS_5$
corresponds to
 \emph{open string}
with infinite extend. Then in
order to derive finite value of the
string action that is evaluated on the
classical solution we have to impose
cut-off on the time and space extend of
the world-sheet theory. It is
clear that the same cut-off has to be
performed for $S^5$  part of the action
as well. Imposing this cut-off
we can explicitly evaluate the
action on given solutions that presumably
give some interesting phenomena in
dual CFT.

As the second example  we  study
another class of the euclidean solution
that was  given in
\cite{Kruczenski:2007cy} and in
\cite{Roiban:2007ju}. This solution
arises from euclidean continuation of
the world-sheet time coordinate from
the homogeneous solutions
 \cite{Frolov:2003qc,Arutyunov:2003za}.
We find that for this solution
Virasoro constraints corresponding dynamics
on $AdS_5$ are non-zero and
consequently the solution on $AdS_5$ is
related to the solution on $S^5$.
This result suggests that
 an equivalence
between euclidean continuation of the
homogeneous solutions given in
\cite{Frolov:2003qc,Arutyunov:2003za}
and light-like Wilson loop solution-where
dynamics on $AdS_5$ decouples from
dynamics on $S^5$-does not  generally
hold when we include
non-trivial dynamics on $S^5$.


The organization of this paper is as
follows. In the next section
(\ref{second}) we introduce the
notation and we solve the equations of
motion for the string with euclidean
world-sheet theory that moves on $S^5$.
We analyze two particular solutions,
the first one corresponding to the
homogeneous motion and the second one
that is analogue of the magnon solution
given in \cite{Hofman:2006xt}.
 Finally
in conclusion (\ref{fourth}) we outline
our results and suggest possible extensions
 of this work.

\section{Euclidean String
on $AdS_5\times S^5$}\label{second}
 Our goal is to
study exact solutions of the closed
string theory in $AdS_5\times S^5$
where the fundamental string has
world-sheet theory with the euclidean
metric signature while the target
space-time has  Minkowski signature.
Before we start this analysis we review
notations for coordinates in $AdS_5$
\cite{Kruczenski:2007cy}. In global
coordinates
$(\rho,t,\phi,\theta_1,\theta_2)$ the
line element of  $AdS_5$ takes the form
\begin{equation}
ds^2=R^2(d\rho^2-\cosh^2\rho
dt^2+\sinh^2\rho
(d\phi^2+\cos^2\phi_1d\theta_1^2
+\sin^2\phi d\theta_2^2)) \ ,
\end{equation}
where $R$ is radius of $AdS_5$ and
$S^5$. It is convenient to introduce
the embedding coordinates $X^M , \
M=(0,\dots,5)$ on which $SO(4,2)$ is
acting linearly.
In these coordinates
 the line element takes the form
\begin{eqnarray}
ds^2&=&dX^M\eta_{MN} dX^N \ ,
\quad \eta_{MN}=(-1,1,1,1,1,-1) \ .
\nonumber \\
\end{eqnarray}
Note that the global coordinates  are
related to the embedding coordinates as
\begin{equation}
X^0+iX^5=R\cosh \rho e^{it} \ , \quad
X^1+iX^2=R\sinh \rho \cos\phi
e^{i\theta_1} \ , \quad
X^3+iX^4=R\sinh\rho \sin\phi
e^{i\theta_2}  \ .
\end{equation}
We can also
 introduce  Poincare coordinates where the
 boundary of $AdS_5$ is at
 $z=0$
\begin{equation}
ds^2=\frac{1}{z^2}(dx^mdx_m+dz^2) \ ,
\quad x^mx_m=-x_0^2+x_i^2 \ , \quad
i=1,2,3\ ,
\end{equation}
where
\begin{eqnarray}
X^0=\frac{x_0}{z} \ , \quad
X^i=\frac{x_i}{z} \ , \quad
X^4=\frac{1}{2z}(-1+z^2+x^mx_m) \ ,
\quad X^5=\frac{1}{2z}(1+z^2+x^mx_m) \
.
\nonumber \\
\end{eqnarray}
Finally, we also use the embedding
coordinates $Y^P \ , P=0,\dots,5$ for
the embedding coordinates of $S^5$ with
euclidean metric $\delta_{PQ}$. Then
the dynamics of bosonic string on
$AdS_5\times S^5$  is governed by the
action
\begin{eqnarray}\label{Sg}
S&=&S^{AdS_5}+S^{S_5} \ , \nonumber \\
S^{AdS_5}&=& \frac{1}{4\pi\alpha'} \int
d\sigma d\tau \sqrt{\gamma}
(\gamma^{\alpha\beta}\partial_\alpha
X^M
\partial_\beta X^N\eta_{MN}+\Lambda (X^M\eta_{MN}X^N+R^2)) \ ,
\nonumber \\
S^{S_5}&=&\frac{1}{4\pi\alpha'} \int d\tau
d\sigma \sqrt{\gamma}
(\gamma^{\alpha\beta}\partial_\alpha
Y^P\partial_\beta
Y^Q\delta_{PQ}+\tLambda
(Y^P\delta_{PQ}Y^Q-R^2)) \ ,
\end{eqnarray}
where $\gamma_{\alpha\beta}$ is
world-sheet metric with euclidean
signature so that in the conformal
gauge the line element takes the form
$ds^2=d\tau^2+ d\sigma^2$.

Finally, $\Lambda, \tLambda$
given in (\ref{Sg})
 are Lagrange multipliers that
 impose conditions $
 X^M\eta_{MN}X^N=-R^2 \ , Y^P\delta_{PQ}
Y^Q=R^2$.
Using (\ref{Sg})  it is
easy to determine corresponding
equations of motion. The variation of
$S^{AdS_5}$ with respect to $X^M$ and
with respect to $\Lambda$ gives
\begin{eqnarray}\label{eqXM}
& &\partial_\alpha (\sqrt{\gamma}
\gamma^{\alpha\beta}\partial_\beta X^M)
-\Lambda X^M=0 \ , \nonumber \\
& &X^MX_M=-R^2 \ .  \nonumber \\
\end{eqnarray}
If we multiply the first equation
in (\ref{eqXM}) with $X_M$ we obtain
\begin{eqnarray}\label{eqXMl}
\Lambda=\frac{1}{R^2}\partial^a
X^M\partial_a X_M \ ,
\nonumber \\
\end{eqnarray}
where we also used the conformal gauge
$\gamma_{\alpha\beta}=\delta_{\alpha\beta}$.
In the same way the variation of $S^{S_5}$
 with respect to $Y^P$
and $\tLambda$ gives
\begin{eqnarray}\label{eqY}
& &\partial_\alpha [
\delta^{\alpha\beta}
\partial_\beta Y^P]
-\tLambda Y^P=0 \ , \nonumber \\
& &Y^P\delta_{PQ}Y^Q=R^2 \ , \quad
\tLambda=-\frac{1}{R^2}\partial_\alpha
Y^P\delta_{PQ}\partial_\beta
Y^Q\delta^{\alpha\beta} \ .
\nonumber \\
\end{eqnarray}
 Following \cite{Kruczenski:2007cy}
we consider
an ansatz that
for $z=0$ ends on two light-like lines
at boundary:
\begin{equation}\label{zans}
z=\sqrt{2}u \ , \quad
u=e^{\alpha\tau-\beta\sigma} \ , \quad
\chi=\alpha\sigma+\beta\tau \ ,
\end{equation}
where $\alpha,\beta$ are real
parameters.
If we write the line
element in Poincare coordinates
as
\begin{equation}
ds^2=\frac{1}{z^2}
(dz^2-du^2+u^2d\chi^2+dx_2^2+dx_3^2) \
, \quad  x_0=u\cosh\chi \ , \quad  x_1=u\sinh\chi \
\end{equation}
then the ansatz (\ref{zans}) takes the form
\begin{equation}
x_0=e^{\alpha\tau-\beta\sigma}\cosh(\alpha
\sigma+\beta\tau) \ , \quad
x_1=e^{\alpha\tau-\beta\sigma}
\sinh(\alpha\sigma+\beta\tau)  \ , \quad
x_2=x_3=0 \ .
\end{equation}
Finally, in the embedding coordinates
the ansatz (\ref{zans}) takes the form
\begin{eqnarray}\label{anstAd}
X^0&=&\frac{R}{\sqrt{2}} \cosh
(\alpha\sigma+\beta\tau) \ , \quad
X^5=\frac{R}{\sqrt{2}}
\cosh(\alpha\tau-\beta\sigma) \ ,
\nonumber \\
X^1&=&\frac{R}{\sqrt{2}}\sinh
(\alpha\sigma+\beta\tau) \ , \quad
X^4=\frac{R}{\sqrt{2}}\sinh(\alpha\tau-\beta\sigma)
\ , \quad  X^2=X^3=0 \ .
\nonumber \\
\end{eqnarray}
With analogy with (\ref{anstAd}) we
propose following  ansatz for the
motion of string on $S^5$
\begin{eqnarray}\label{SY}
Y^1&=&\frac{R}{\sqrt{2}}
\cos(\gamma\tau+\delta\sigma) \ , \quad
Y^2=\frac{R}{\sqrt{2}}\sin
(\gamma\tau+\delta\sigma) \ ,
\nonumber \\
Y^3&=&\frac{R}{\sqrt{2}}
\cos(\gamma\sigma-\delta\tau) \ , \quad
Y^4=\frac{R}{\sqrt{2}}
\sin(\gamma\sigma-\delta\tau) \ ,
\nonumber \\
\end{eqnarray}
where $\gamma,\delta$ are constants.
 Now it is easy to see that for
the ansatz (\ref{anstAd})
$\Lambda$ takes the form
\begin{eqnarray}
\Lambda=
(\alpha^2+\beta^2) \nonumber \\
\end{eqnarray}
and consequently the ansatz
(\ref{anstAd}) solves the equation of
motion (\ref{eqXM}). In the same way
(\ref{SY}) gives
\begin{eqnarray}
\tLambda=-(\gamma^2+\delta^2)  \
\end{eqnarray}
and again
 it is easy to see that the
equation of motion (\ref{eqY})
 are satisfied as well.

As the next step we impose Virasoro
constraints.  Explicitly, the variation
of the action (\ref{Sg}) with respect
to world-sheet metric
$\gamma^{\alpha\beta}$  implies the
constraints
\begin{equation}
T_{\alpha\beta}=\frac{2\pi}{\sqrt{\gamma}}
\frac{\delta S}{\delta
\gamma^{\alpha\beta}}=
T_{\alpha\beta}^{AdS_5}+
T_{\alpha\beta}^{S_5}= 0 \ ,
\end{equation}
where
\begin{eqnarray}
T_{\tau\tau}^{AdS_5}
&=&-T^{AdS_5}_{\sigma\sigma}=\frac{1}{2\alpha'}
(\partial_\tau X^M\partial_\tau X_M-
\partial_\sigma X^M\partial_\sigma X_M)
\ ,
\nonumber \\
T^{AdS_5}_{\tau\sigma}&=&\frac{1}{\alpha'}\partial_\tau
X^M
\partial_\sigma X_M \ ,
\nonumber \\
T_{\tau\tau}^{S_5}&=&-
T^{S_5}_{\sigma\sigma}=
\frac{1}{2\alpha'}(\partial_\tau
Y^P\partial_\tau Y_P-\partial_\sigma
Y^P\partial_\sigma Y_P) \ , \nonumber \\
T^{S_5}_{\tau\sigma}&=&\frac{1}{\alpha'}\partial_\tau
Y^P\partial_\sigma Y_P \ ,
\nonumber \\
\end{eqnarray}
where we considered the metric in the form
$\gamma_{\alpha\beta}=\mathrm{diag}(1,1)$
and in the final step we  used the equation of
motion for $\Lambda$ and $\tLambda$.
 Now for the ansatz
(\ref{anstAd})  we obtain
\begin{eqnarray}
T_{\tau\tau}^{AdS_5}=0 \ , \quad
T_{\tau\sigma}^{AdS_5}=0 \ .
\nonumber \\
\end{eqnarray}
 On the other hand the
$S^5$ part of the Virasoro constraints imply
\begin{eqnarray}
T^{S_5}_{\tau\tau}&=&
\frac{1}{4\alpha'}(\gamma^2-\delta^2)=0
 \ , \quad
T_{\tau\sigma}^{S_5}=0 \ . \nonumber \\
\nonumber \\
\end{eqnarray}
The first constraint implies that $
\gamma^2=\delta^2$ while the second one
is automatically satisfied. However the
analysis performed above suggests that
the dynamics on $AdS_5$ decouples from
the dynamics on $S^5$ for ansatz
(\ref{anstAd}). On the other hand when
we evaluate action on given solutions
we have to take integration cut-off
into account.  In fact note
 that
the  $AdS_5$ part of the action is
equal to
\begin{equation}\label{Sev}
S^{AdS_5}=\frac{\sqrt{\lambda}}{4\pi}(\alpha^2+\beta^2)
\int d\tau d\sigma \ ,
\end{equation}
where
$\sqrt{\lambda}=\frac{R^2}{\alpha'}$.
As in \cite{Kruczenski:2007cy} we
introduce following cut-off
prescription
\begin{equation}\label{cutoff}
\ln l<\alpha\tau-\beta\sigma <\ln L \ ,
\quad -\frac{\Gamma}{2}<
\alpha\sigma+\beta\tau<\frac{\Gamma}{2} \ .
\end{equation}
As the next step we introduce
the  coordinates
\begin{eqnarray}
m&=&\alpha\tau-\beta\sigma \ , \quad
n=\alpha\sigma+\beta\tau \ , \nonumber \\
\tau &=&\frac{\alpha m+\beta n}
{\alpha^2+\beta^2} \ , \quad
\sigma=\frac{n\alpha-\beta
m}{\alpha^2+\beta^2} \ ,  \nonumber \\
\end{eqnarray}
where the Jacobian of the
transformation from $(\tau,\sigma)$ to
$(m,n)$ is equal to
$J=\frac{1}{\alpha^2+\beta^2}$. Then we
can easily evaluate (\ref{Sev}) and we
obtain
\begin{eqnarray}
S^{AdS_5}&=&
=\frac{\sqrt{\lambda}}{4\pi}(\alpha^2+\beta^2)
\int d\tau d\sigma=\nonumber \\
&=&\frac{\sqrt{\lambda}}{4\pi}(\alpha^2+\beta^2)
\int_{\ln l}^{\ln L} dm
\int_{-\frac{\Gamma}{2}}^{\frac{\Gamma}{2}}
dn J=\nonumber \\
&=&
\frac{\sqrt{\lambda}}{4\pi}\Gamma\ln
 \frac{L}{l} \ .
 \nonumber \\
\end{eqnarray}
We see that the value of the action
does not depend on $\alpha,\beta$. On
the other hand when we evaluate
 $S^{S_5}$ for the ansatz (\ref{SY})
we obtain
\begin{eqnarray}
S^{S_5}=\frac{\sqrt{\lambda}}{4\pi}
\int d\tau d\sigma (\gamma^2+\delta^2)=
\frac{\sqrt{\lambda}}{4\pi}\frac{\gamma^2+\delta^2}
{\alpha^2+\beta^2}\Gamma \ln
\frac{L}{l} \ .   \nonumber \\
\end{eqnarray}
However it would be more natural to
express given action in terms of
conserved charges. In fact, it is easy
to see that the action $S^{S_5}$ is
manifestly invariant under rotation
\begin{equation}
Y'^M=\Omega^M_NY^N\approx Y^M+
\omega^M_NY^N \ , \quad
\omega^M_N=-\omega^N_M \ll 1
\end{equation}
that implies an existence of
following conserved
charges
\begin{equation}\label{defJ}
J^{MN}=\frac{1}{4\pi\alpha'} \int
d\sigma (Y^M\partial_\tau Y^N-
Y^N\partial_\tau Y^M) \ .
\end{equation}
Let us now define
\begin{equation}\label{defJ1}
J_1=J^{12} \ , \quad J_2=J^{34} \ ,
\quad J_3=J^{56} \ .
\end{equation}
Then for (\ref{SY})  we obtain
\begin{eqnarray}
J_1=\frac{\sqrt{\lambda}}{8\pi} \gamma
\int d\sigma \ , \quad  J_2=
-\frac{\sqrt{\lambda}}{8\pi}
\gamma \int d\sigma \ .  \nonumber \\
\end{eqnarray}
It turns out, however that
for integration domain defined in
(\ref{cutoff}) these charges
 explicitly depend on time.
For that reason we restrict ourselves
to the case when $\beta=0$. Then
$-\frac{\Gamma}{2\alpha}<\sigma<\frac{\Gamma}{2\alpha}$
 and we obtain
\begin{eqnarray}
J_1&=&-J_2=\frac{\sqrt{\lambda}} {8\pi}
\gamma \int_{-\frac{\Gamma}{2\alpha}}^
{\frac{\Gamma}{2\alpha}}d\sigma
=\frac{\sqrt{\lambda}}{8\pi}
\frac{\gamma}{\alpha} \Gamma \
\nonumber \\
\end{eqnarray}
and hence we can write
\begin{equation}
S^{S_5}=32 \frac{\pi}{\sqrt{\lambda}}
J_1^2\frac{1}{\Gamma}\ln \frac{L}{l}=
16\frac{\pi}{\sqrt{\lambda}}
\frac{1}{\Gamma}\ln \frac{L}{l}
(J_1^2+J_2^2) \ .
\end{equation}
In summary we obtain that the action
evaluated on the solution is equal to
\begin{equation}
S=\frac{\sqrt{\lambda}}{4\pi}\ln
\frac{L}{l}\Gamma+ 16\frac{\pi}{
\sqrt{\lambda}} \frac{1}{\Gamma}\ln
\frac{L}{l} (J_1^2+J_2^2) \ .
\end{equation}
To conclude, we  found generalization
of the solution
\cite{Kruczenski:2007cy} where we
included non-trivial dynamics on $S^5$.
We also shown that this solution is
valid for any values of parameters
$\alpha,\beta$ and that  Virasoro
constraints do not imply any relation
between the motion on $AdS_5$ and
$S^5$. On the other hand we have argued
that when we wanted to evaluate
world-sheet action on these solutions
we had to impose the same integration
cut-off in both parts of the action.

Let us now consider  ansatz that is an
analogue of the magnon-like solution
\cite{Hofman:2006xt}. To do this we
restrict ourselves to the motion of the
string on $S^2$. Then it is convenient
to use the parametrization
\begin{equation}
Y^1=R\sin\theta \cos\phi \ , \quad Y^2=
R\sin\theta\sin\phi \ , \quad Y^3=
R\cos\theta \
\end{equation}
so that the line element on two sphere
$S^2$  takes the form
\begin{equation}
ds^2=R^2(d\theta^2+\sin^2\theta
d\phi^2) \ .
\end{equation}
Then  the action that determines
motion on $S^2$ takes the form
\begin{equation}\label{S2}
S^{S_2}=\frac{R^2}{4\pi\alpha'}\int d\tau
d\sigma [\delta^{\alpha\beta}
\partial_\alpha
\theta\partial_\beta\theta+
\sin^2\theta \delta^{\alpha\beta}
\partial_\alpha \phi\partial_\beta\phi]
\ .
\end{equation}
Further we  presume that the
motion on $AdS_5$ is determined
by the ansatz
(\ref{anstAd}) while for the motion on
$S^2$ we propose an ansatz
\begin{equation}\label{ansMald}
\theta=\theta(y) \ , \quad
\phi=\omega\tau+\tphi(y) \ ,
\end{equation}
where
\begin{equation}
y=\gamma\tau+\delta\sigma \ .
\end{equation}
Using the simple form of the action
(\ref{S2}) it is  easy to determine
corresponding equations of motion for
$\phi$
\begin{eqnarray}\label{eqphi}
 [\sin^2\theta
((\gamma^2+\delta^2)
\tphi'+\omega\gamma)]'=0
\end{eqnarray}
while the equation of motion  for
 $\theta$ takes
the form
\begin{equation}
(\gamma^2+\delta^2)\theta''
-\sin\theta\cos\theta ((\omega+
\gamma\tphi')^2+\delta^2\tphi'^2)=0 \ ,
\end{equation}
where
$(\dots)'\equiv\frac{d(\dots)}{dy}$.
Note that (\ref{eqphi}) implies
\begin{equation}\label{tphi1}
\tphi'=\frac{1}{\gamma^2+\delta^2}
[\frac{B}{R^2\sin^2\theta}-\omega\gamma]
 \ ,
\end{equation}
where $B$ is constant. Further, using
the fact that $T_{\alpha\beta}^{AdS_5}$
vanishes separately we  use the
Virasoro constraints
$T_{\tau\tau}^{S_2}=0$ in order  to determine
differential equation for $\theta$
\begin{eqnarray}\label{ttheta}
\theta'^2&=&
-\frac{\sin^2\theta}{\gamma^2-\delta^2}
[(\omega+\gamma\tphi')^2-\delta^2
\tphi'^2]= \nonumber \\
&=&-\frac{1}{(\gamma^2+
\delta^2)^2}[\frac{
B^2}{R^4\sin^2\theta}-\omega^2\delta^2\sin^2\theta
+\frac{4\gamma B\omega \delta^2}{R^2
(\gamma^2-\delta^2)}] \nonumber \\
\end{eqnarray}
using (\ref{tphi1}).
On the other hand when we consider the
second Virasoro constraint $T_{\tau\sigma}^{S_2}=0$
we obtain
\begin{equation}
T^{S_2}_{\tau\sigma}
=\theta'^2\gamma\delta+\sin^2\theta
(\omega\delta \tphi'+\gamma\delta
\tphi'^2)=0 \
\end{equation}
that together with the constraint
$T^{S_2}_{\tau\tau}=0$ implies
\begin{eqnarray}\label{Ttsigma}
\tphi'=-\frac{\omega\gamma}{\gamma^2+\delta^2}\
\end{eqnarray}
and consequently
\begin{equation}
\phi=\frac{\omega\delta}{\gamma^2+\delta^2}
(\delta\tau-\gamma\sigma) \ .
\end{equation}
Further, when we compare
(\ref{Ttsigma}) with (\ref{tphi1}) we
obtain that $B=0$ and consequently
(\ref{ttheta}) implies following
differential equation
\begin{equation}
\theta'=\frac{\omega\delta}
{\gamma^2+\delta^2}\sin\theta
\end{equation}
that has the solution
\begin{equation}
\cos\theta=-\frac{\sinh(\frac{\omega\delta}{
\gamma^2+\delta^2}(\gamma\tau+\delta\sigma))}
{\cosh
(\frac{\omega\delta}{\gamma^2+\delta^2}
(\gamma\tau+\delta\sigma))} \ .
\end{equation}

 Let us now evaluate the
 $S_2$ part of the action
   for the ansatz (\ref{ansMald})
\begin{eqnarray}\label{S2ev}
S^{S_2}
&=&\frac{\sqrt{\lambda}}{4\pi} \int
d\sigma d\tau
[\theta'^2(\gamma^2+\delta^2)+
\sin^2\theta
[(\omega+\gamma\tphi')^2+\delta^2\tphi'^2]]=
\nonumber \\
&=&\frac{\sqrt{\lambda}}{2\pi} \int
d\tau d\sigma
\frac{\omega^2\delta^2}{\gamma^2+\delta^2}
\sin^2\theta \ . \nonumber \\
\end{eqnarray}
In order to evaluate
the action (\ref{S2ev}) appropriately
we have to impose the integration
cut-off that arise from the analysis
of the dynamics of  $AdS_5$
string and we obtain
\begin{eqnarray}
S^{S_2}
&=&\frac{\sqrt{\lambda}}{2\pi}
\frac{\omega^2\delta^2}
{\gamma^2+\delta^2}\times \nonumber \\
&\times &
 \int_{\frac{1}{\alpha\ln l}}^{\frac{1}{\alpha}
\ln L} d\tau
\int_{-\frac{\Gamma}{2\alpha}}
^{\frac{\Gamma}{2\alpha}} d\sigma
\frac{1}{ \cosh^2 (\frac{\omega\delta}
{\gamma^2+\delta^2}(\gamma\tau+\delta\sigma))}=
\frac{\sqrt{\lambda}}
{\pi\alpha}\ln\frac{L}{l} \ .
\nonumber \\
\end{eqnarray}
Let us calculate the charge related to the isometry
along  $\phi$ direction
\begin{eqnarray}
J_\phi&=&\frac{\sqrt{\lambda}}{2\pi}
\int_{-\frac{\Gamma}{2\alpha}}^
{\frac{\Gamma}{2\alpha}} d\sigma
\sin^2\theta (\omega+\gamma\phi')=
\nonumber \\
&=& \frac{\sqrt{\lambda}}{2\pi}
\frac{\delta^2\omega}{\gamma^2+\delta^2}
\int_{-\frac{\Gamma}{2\alpha}}^
{\frac{\Gamma}{2\alpha}} d\sigma
\frac{1}{ \cosh^2 (\frac{\omega\delta}
{\gamma^2+\delta^2}(\gamma\tau+\delta\sigma))}=
\frac{\sqrt{\lambda}}{2\pi\alpha}
\nonumber \\
\end{eqnarray}
and hence we obtain the result that
\begin{equation}
S^{S_2}=2J_\phi\ln\frac{L}{l} \ .
\end{equation}
Interestingly, due to the profile
of the classical solution the action
does not depend on the spatial cut-off
$\Gamma$. It would be certainly
interesting to  find the dual CFT interpretation
of such a configuration.

Now we proceed to the generalization of
the second type of the euclidean
world-sheet solution
\cite{Roiban:2007ju}. We start with the
equations of motion for bosonic string
with Minkowski metric
\begin{eqnarray}\label{eqXMm}
& &\partial_\alpha [ \eta^{\alpha\beta}
\partial_\beta X^M]
-\frac{1}{R^2}\left(\eta^{\alpha\beta}
\partial_{\alpha} X^N
\partial_{\beta}
X_N\right)X^M=0 \ , \nonumber
\\
& &\partial_\alpha [\eta^{\alpha\beta}
\partial_\beta Y^P]
+\frac{1}{R^2}\left(\eta^{\alpha\beta}\partial_\alpha
Y^Q\partial_\beta Y_Q\right)
Y^P=0 \ ,
\nonumber \\
\end{eqnarray}
where $\eta=\mathrm{diag}(-1,1)$.
Let us then consider following ansatz
\footnote{The solution (\ref{X0M})
belongs to the class of homogeneous
string solutions as the rigid circular
string found in
\cite{Frolov:2003qc,Arutyunov:2003za}.}
\begin{eqnarray}\label{X0M}
X^0&=&R \cosh\talpha\sigma
\cos\tbeta\tau' \ , \quad X^1=R
\sinh\talpha\sigma\cos\tbeta\tau' \ ,
\nonumber \\
X^5&=&R \cosh\talpha\sigma \sin\tbeta
\tau' \ , \quad
X^2=R\sinh\talpha\sigma\sin\tbeta\tau'
\ , \nonumber \\
\end{eqnarray}
where $\tau'$ is now time-coordinate
on Minkowski world-sheet.
We again find that the equations of
motion are satisfied for any
$\talpha,\tbeta$.
As the next step
 we perform an analytic
continution $\tau'=-i\tau$ and hence
\begin{eqnarray}
X^0=R \cosh\talpha\sigma
\cosh\tbeta\tau \ , \quad X^1=R
\sinh\talpha\sigma\cosh\tbeta\tau \ ,
\nonumber \\
X^5=iR \cosh\talpha\sigma \sinh\tbeta
\tau \ , \quad
X^2=iR\sinh\talpha\sigma\sinh\tbeta\tau'
\ . \nonumber \\
\end{eqnarray}
If we now write
 $X^2=iX'^5 \ ,
X^5=iX'^2$ we obtain
\begin{eqnarray}
X'^0&=&R \cosh\talpha\sigma
\cosh\tbeta\tau \ , \quad X'^1=R
\sinh\talpha\sigma\cosh\tbeta\tau \ ,
\nonumber \\
X'^2&=&R \cosh\talpha\sigma \sinh\tbeta
\tau \ , \quad
X'^5=R\sinh\talpha\sigma\sinh\tbeta\tau
\ . \nonumber \\
\end{eqnarray}
Finally we perform rotation in $(0,5)$
and $(1,2)$ plane and we obtain
\begin{eqnarray}\label{Tm}
X^0&=&\frac{X'^0+X'^5}{\sqrt{2}}=
\frac{R}{\sqrt{2}}\cosh
(\talpha\sigma+\tbeta\tau) \ , \nonumber \\
X^5&=&\frac{X'^0-X'^5}{\sqrt{2}}=
\frac{R}{\sqrt{2}} \cosh
(\talpha\sigma-\tbeta\tau) \ , \nonumber \\
X^1&=&\frac{X'^1+X'^2}{\sqrt{2}}=
\frac{R}{\sqrt{2}}\sinh
(\talpha\sigma+\tbeta\tau) \ ,
\nonumber \\
X^2&=&\frac{X'^1-X'^2}{\sqrt{2}}=
\frac{R}{2}\sinh(\talpha\sigma-\tbeta\tau)
\ . \nonumber \\
\end{eqnarray}
Let us now check properties of the
ansatz (\ref{Tm}). Firstly, it is easy
to see that $\Lambda$ is equal to
\begin{equation}
\Lambda=(\talpha^2+\tbeta^2) \
\end{equation}
and hence the equations of motions are
satisfied. On the other hand the
Virasoro constraints
$T^{AdS_5}_{\tau\tau}=-T^{AdS_5}_{\sigma\sigma}=0$
is equal to
\begin{eqnarray}
T^{AdS_5}_{\tau\tau}=\frac{\sqrt{\lambda}}
{4} (\tbeta^2-\talpha^2)
\end{eqnarray}
and we see that $AdS_5$ part of the
stress energy tensor does not vanish.
Then let us again consider the motion
on $S^5$ that is parameterized with the
ansatz (\ref{SY}). In this case
Virasoro constraints $T_{\tau\tau}=
T_{\tau\tau}^{AdS_5}+T_{\tau\tau}^{S_5}=0$
implies
\begin{equation}\label{Virm}
\tbeta^2-\talpha^2+\gamma^2-\delta^2=0
\ .
\end{equation}
Consequently the action evaluated on
the ansatz (\ref{SY}) and (\ref{Tm}) is
equal to
\begin{eqnarray}
S&=&\frac{\sqrt{\lambda}} {4\pi}
[\tbeta^2+\talpha^2+\gamma^2+\delta^2]
\int d\tau d\sigma= \nonumber \\
&=&\frac{\sqrt{\lambda}}{2\pi\tbeta^2}
[\tbeta^2+\gamma^2]\Gamma \ln
\frac{L}{l} \ ,
\nonumber \\
\end{eqnarray}
where now we chose the integration
cut-off
$-\frac{\Gamma}{2\tbeta}<\sigma<\frac{\Gamma}{2\tbeta}
\ , \frac{1}{\tbeta}\ln l <\tau<
\frac{1}{\tbeta} \ln L$. Then  we
introduce charges
\cite{Tseytlin:2003ii}
\begin{eqnarray}
S_0=\frac{1}{2\pi\alpha'} \int d\sigma
(X^5 \partial_\tau X^0-X^0\partial_\tau
X^5) \ , \quad
S_1=\frac{1}{2\pi\alpha'} \int d\sigma
(X^1 \partial_\tau
X^2-X^2\partial_\tau X^1) \nonumber \\
\end{eqnarray}
that for (\ref{Tm}) are equal to
\begin{eqnarray}
S_0&=&\tbeta\frac{\sqrt{\lambda}}
{4\pi} \int d\sigma \sinh
2\talpha\sigma
=\nonumber \\
&=& \frac{\sqrt{\lambda}}{4\pi}
\frac{\tbeta}{\talpha} \cosh
\frac{\talpha}{\tbeta} \Gamma\approx
\frac{\sqrt{\lambda}}
{4\pi}\frac{\tbeta}{\talpha}e^{\frac{\talpha}{\tbeta}
\Gamma}
 \ , \quad
S_1=-S_0 \ . \nonumber \\
\end{eqnarray}
In the same way we find that $J_1,J_2$
defined in (\ref{defJ}) and (\ref{defJ1})
are equal to
\begin{equation}
J_1=\frac{\sqrt{\lambda}} {8\pi} \frac{
\gamma} {\tbeta} \Gamma \ , \quad
J_2=-\frac{\sqrt{\lambda}} {8\pi}
\frac{ \delta}{\tbeta} \Gamma \ .
\end{equation}
 Then we can write the action in an
alternative form
\begin{equation}
S=\frac{8\pi}{\sqrt{\lambda}} S_0^2
e^{-2\frac{\talpha}{\tbeta}
\Gamma}\Gamma
\ln \frac{L}{l}+
\frac{32\pi}{\sqrt{\lambda}
}\frac{J_1^2}{\Gamma}
 \ln \frac{L}{l} \ .
\end{equation}
It is important to stress that now
$\talpha$ and $\tbeta$ are not
arbitrary but are determined by charges
$J_1,J_2$ through the relation
(\ref{Virm}) . In particular, the
condition $\talpha=\tbeta=1$ can be
imposed in case when
$\gamma^2-\delta^2=0$ (equivalently
when $J_1^2=J_2^2$). We will discuss
consequence of this result below.

 Let us
again consider magnon-like solution
where now we have to take into account
nonzero contribution from $AdS_5$
part of the Virasoro constraint that we
denote as
 $T_{AdS_5}\equiv \frac{1}{4\alpha'} \kappa^2$. Then the
 vanishing of the total $T_{\tau\tau}=
 T^{AdS_5}_{\tau\tau}+T^{S^2}_{\tau\tau}=0$
 implies
\begin{eqnarray}\label{tthetak}
\theta'^2
&=&-\frac{\kappa^2}{R^2(\gamma^2-\delta^2)}
-\frac{\sin^2\theta}{\gamma^2-\delta^2}
[(\omega+\gamma\tphi')^2-\delta^2
\tphi'^2]= \nonumber \\
&=&-\frac{1}{(\gamma^2+
\delta^2)^2}[\frac{
B^2}{R^4\sin^2\theta}-\omega^2\delta^2\sin^2\theta
+\frac{\kappa^2}{R^2}(\gamma^2-\delta^2)]
\ .
\nonumber \\
\end{eqnarray}
On the other hand  the
second Virasoro constraint $T_{\tau\sigma}=0$
implies
\begin{equation}
T_{\tau\sigma}
=\theta'^2\gamma\delta+\sin^2\theta
(\omega\delta \tphi'+\gamma\delta
\tphi'^2)=0 \
\end{equation}
that together with the constraint
$T_{\tau\tau}=0$ gives
\begin{eqnarray}\label{Ttsigmam}
\kappa^2\gamma+B\omega=0 \nonumber \\
\end{eqnarray}
using also the fact that the equation
of motion for $\phi$ takes the form
\begin{equation}\label{tphi1b}
\tphi'=\frac{1}{\gamma^2+\delta^2}
[\frac{B}{R^2\sin^2\theta}-\omega\gamma]
 \ .
\end{equation}
Finally we obtain
\begin{eqnarray}
\theta'^2&=&\frac{\omega^2\delta^2} {
(\gamma^2+\delta^2)^2\sin^2\theta}
[(\sin^2\theta-M)( \sin^2\theta-N)]
\nonumber \ ,
\\
M&=&-\frac{\kappa^2}{\omega^2R^2} \ ,
\quad  N=\frac{\kappa^2\gamma^2}
{\delta^2 R^2\omega^2} \ . \nonumber \\
\end{eqnarray}
We see that in order to have
real solution we should
perform an analytic continuation
\begin{equation}
\omega=i\tomega \ , \gamma=i\tgamma \ .
\end{equation}
Then we finally obtain
\begin{eqnarray}\label{thetadm}
\theta'=\frac{\tomega\delta}{(\delta^2-\tgamma^2)
\sin\theta}\sqrt{(\sin^2\theta-\sin^2\theta_1)
(\sin^2\theta_2-\sin^2\theta)} \ ,
\end{eqnarray}
where
\begin{equation}
\sin^2\theta_1=\frac{\kappa^2}{\tomega^2
R^2} \ , \quad \sin^2\theta_2=
\frac{\kappa^2 \tgamma^2}{R^2 \delta^2
\tomega^2} \ .
\end{equation}
Following \cite{Bobev:2007bm}
we can  now distinguish
two limiting configurations: either giant
magnon solution where
$\sin^2\theta_1=1$ or spike solution
where $\sin^2\theta_2=1$. Let us
firstly consider giant magnon solution.
The action evaluated on this solution
takes the form
\begin{eqnarray}
S&=&\frac{\sqrt{\lambda}}{4\pi} \int
d\tau d\sigma
[(\delta^2-\tgamma^2)\theta'^2+
\sin^2\theta
\frac{\tomega^2\delta^2}{\delta^2-\tgamma^2}
+\frac{B^2}{(\delta^2-\tgamma^2)R^4\sin^2\theta}]
=\nonumber \\
&=&\frac{\tomega ^2\delta ^2
\sqrt{\lambda}}{4\pi(\delta^2-\tgamma^2)}
\int d\tau d\sigma
\frac{\sin^2\theta(1+\sin^2\theta_2)}{\sin^2\theta}=
\frac{\sqrt{\lambda}}{2\pi}\frac{\tomega^2
(\delta^2+\tgamma^2)}
{(\delta^2-\tgamma^2)}
\frac{\Gamma}{\tbeta} \ln \frac{L}{l} \
.
\nonumber \\
\end{eqnarray}
Interestingly, for spike solution
($\sin^2\theta_2=
\frac{\kappa^2 \tgamma^2}{R^2 \delta^2
\tomega^2}=1$) we find
\begin{equation}
S=\frac{\sqrt{\lambda}}{2\pi}
\frac{\tomega^2 \delta^2
(\delta^2+\tgamma^2) } {
\tgamma^2(\delta^2-\tgamma^2)}
\frac{\Gamma}{\tbeta} \ln \frac{L}{l} \
,
\end{equation}
where now $\tomega$ is related to
$\kappa^2=R^2(\tbeta^2-\talpha^2)$
through the relation (\ref{Ttsigmam}).

It is important to stress  that for
$\alpha=\beta=1$ the ansatz
(\ref{anstAd}) is related to the ansatz
(\ref{Tm}) where $\talpha=\tbeta=1$. In
other words, the scaling limit of the
spinning closed string solution is
equivalent, upon an analytic
continuation to the euclidean
world-sheet combined with a discrete
$SO(2,4)$ rotation in $AdS_5$, to the
global $AdS_5$ version of the null cusp
solution given in (\ref{anstAd}) (For
$\alpha=\beta=\talpha=\tbeta=1$).
However as was shown above in case when
we include non-trivial dynamics on
$S^5$ this is not generally true since
now $\talpha,\tbeta$ are functions of
the conserved charges related to the
dynamics on $S^5$ while in case of the
null cusp solution  (\ref{anstAd}) the
dynamics on $AdS_5$ decouples from the
dynamics on $S^5$ at least on the
classical level.


%

\section{Conclusion}\label{fourth}
In this section we
give short summary of results derived in
this paper.
Our goal was to study some
 solutions of \emph{open string}
 with euclidean world-sheet that propagates
in $AdS_5\times S^5$. We started with
the review of solution presented in
\cite{Kruczenski:2007cy} and we
found that  this solution is
valid for  any real parameters $\alpha,\beta$.
We also shown that Virasoro constraints
for this solution vanish. This has
an important consequence when we included
non-trivial dynamics  on $S^5$ that now
naively decouples from the dynamics on
 $AdS_5$. 

Then we also considered second form of the
solution that was presented in
\cite{Roiban:2007ju}. We shown that
this solution has non-zero contribution
from the $AdS_5$ part and hence
the motion on $S^5$ does not decouple
from the motion on $AdS_5$. We calculated
the values of world-sheet actions on
these solutions.

To conclude we mean that the dynamics of
euclidean string in $AdS_5\times S^5$ has
many interesting properties that
deserves further study. It would be certainly
very interesting to clarify the relation
 of the solutions  found there to
scattering phenomena in
dual QFT living on the boundary of $AdS_5$.
\\
\\
{\bf Acknowledgement}

This work
 was supported  by the Czech Ministry of
Education under Contract No. MSM
0021622409.

\end{document}